\documentclass[journal]{IEEEtran}

\ifCLASSINFOpdf
\else
\fi
\usepackage{array}
\usepackage{booktabs}
\usepackage{mdwmath}
\usepackage{multicol}
\usepackage[cmex10]{amsmath}
\usepackage{amsfonts}
\usepackage{array,color}
\usepackage{booktabs}

\usepackage{graphicx}
\usepackage{caption}
\usepackage{subcaption}
\usepackage{epstopdf}
\usepackage{cite}
\usepackage{algorithm}
\usepackage{algorithmic}
\usepackage{bm}
\usepackage{csquotes}
\usepackage{amssymb}
\usepackage{marvosym}
\usepackage{multirow}
\usepackage{diagbox}
\usepackage{cancel}
\usepackage{booktabs} 

\usepackage{extarrows}
\usepackage{etoolbox}
\makeatletter
\patchcmd{\@begintheorem}{\textit}{\textbf}{}{}
\makeatother

\newtheorem{theorem}{Theorem}

\newtheorem{lemma}{Lemma}

\newtheorem{corollary}{Corollary}

\usepackage[left=0.6in,right=0.6in,bottom=1in,top=0.7in]{geometry}
\begin{document}

\IEEEoverridecommandlockouts
\IEEEpubid{\begin{minipage}[t]{\textwidth}\ \\[10pt]
		\centering\normalsize{978-1-5090-6008-5/17/\$31.00 \copyright 2018 IEEE \\ This is a preprint version. Personal use of this material is permitted.
			However, permission to use this material for any other purposes must be obtained from the IEEE by sending a request to pubs-permissions@ieee.org.}
\end{minipage}}

\title{\mbox{}\vspace{1.5cm}\\
	\textsc{Design and Implementation of 5G e-Health Systems, Technologies, Use Cases and Future Challenges} \vspace{1.5cm}}
\author{Di~Zhang~\IEEEmembership{Senior Member, ~IEEE,} Joel~J.~P.~C.~Rodrigues,~\IEEEmembership{Fellow,~IEEE,}  Yunkai~Zhai, Takuro~Sato,~\IEEEmembership{Life~Fellow,~IEEE}
	\vspace{1.5cm}
	\thanks{Di~Zhang (email: {\tt dr.di.zhang@ieee.org}) is with the School of Information Engineering, Zhengzhou University, China. Joel J. P. C. Rodriguesis (email:{\tt joeljr@ieee.org}) is with the Federal University of Piau\'i (UFPI), Teresina - PI, Brazil, and also with the Instituto de Telecomunica\c{c}\~oes, Portugal. Yunkai Zhai (email: {\tt zhaiyunkai@zzu.edu.cn}) is with the National Telemedicine Center, the National Engineering Laboratory for Internet Medical Systems and Applications, and the School of Management Engineering, Zhengzhou University, Zhengzhou 450001, China. Takuro~Sato (email: {\tt t-sato@waseda.jp}) is with the School of Fundamental Science and Eingineering, Waseda University, Ookubo 3-4-1, Tokyo 169-8555, Japan.}

	\underline{Manuscript accepted by IEEE Communications Magazine on April 26, 2021.}\\
	978-1-5090-6008-5/17/\$31.00 \copyright 2018 IEEE \\ This is a preprint version. Personal use of this material is permitted.
	However, permission to use this material for any other purposes must be obtained from the IEEE by sending a request to pubs-permissions@ieee.org.
}

\date{\today}
\renewcommand{\baselinestretch}{1.2}
\thispagestyle{empty} \maketitle \thispagestyle{empty}
\newpage
\setcounter{page}{1}

\begin{abstract}
	
Fifth generation (5G) aims to connect massive devices with even higher reliability, lower latency and even faster transmission speed, which are vital for implementing the e-health systems. However, the current efforts on 5G e-health systems are still not enough to accomplish its full blueprint. In this article, we first discuss the related technologies from physical layer, upper layer and cross layer perspectives on designing the 5G e-health systems. We afterwards elaborate two use cases according to our implementations, i.e., 5G e-health systems for remote health and 5G e-health systems for Covid-19 pandemic containment. We finally envision the future research trends and challenges of 5G e-health systems.

\end{abstract}

\IEEEpeerreviewmaketitle

\section{Introduction}

The initial e-health system was introduced in 1955 by Cecil Wittson, M. D., to share lectures between University of Nebraska Medical Center (UNMC) and mental hospitals located in different places. In 1959, Albert J. Jutras reported remote diagnostic consultations based on the fluoroscopy images transmitted by coaxial cable. After that, e-health was adopted widely to provide e-diagnosis and e-treatment services. Dominant benefits of e-health attribute to the reduced time and manpower costs. In literature, e-health can be classified into two categories, i.e., outdoor scenario and indoor scenario. However, due to the weakness of prior generations of wireless technologies, such as fourth generation long-term evolution (4G LTE), the ambition of e-health was not fully accomplished.
	
Different from existing generations of wireless technologies, fifth generation (5G) is an ideal solution for e-health's ambition. This is mainly because of the even faster transmission speed (especially the even faster uplink transmission speed) and the ultra-reliable and low latency communications (URLLC) of 5G\cite{dzhangtcom, ibf}. Thanks to the 5G technologies, we may provide 24 hours in-home health monitoring, timely e-diagnosis and e-treatment services, which are of significant importance to the patient with chronic diseases such as cerebral stoke and myocardial infarction\cite{ttwei}. The 5G e-health systems can also play a significant role on containing the acute infectious diseases. This is achieved by the reduced physical contacts, accelerated disease information flow, and enhanced therapy capacity and efficiency.
	
While deploying 5G e-health systems, we first need to identify the requirement of the specific 5G e-health service, and then tailor the necessary 5G technologies to cater to it. Under this premise, an identity management (IdM) framework for 5G e-health systems was proposed to connect different types of devices with different wireless access technologies\cite{dfang}. It was found that the IdM framework can achieve the required security properties efficiently in the e-health systems. In \cite{mme}, cloud computing-based Internet of things (C-IoT) was introduced to e-health systems, wherein the authors further investigated the energy efficiency performances of various proposals. However, as is known, existing studies about 5G e-health systems are mostly constrained to the network and application layers, perspective from the physical layer is limited\cite{brito}. 

In this article, we first investigate the 5G technologies on designing the 5G e-health systems from both physical layer and upper layers. We afterwards introduce two use cases according to our implementations, i.e., 5G e-health systems for remote health and 5G e-health systems for Covid-19 pandemic containment. On this basis, we exploit a series of field trials to compare the quality of service (QoS) and quality of experience (QoE) performances among the 4G LTE, 5G non-standalone (NSA) and 5G standalone (SA) deployments. The noticeable results about 5G from our field trials are: 1) The current 5G uplink transmission speed is about 100 Mbps, which is almost enough to transmit the high-definition medical streaming data; 2) The current 5G latency is about 10 ms, which is far away from the claimed less than 1 ms latency. We finally discuss the challenging issues on the worldwide deployment of 5G e-health systems. This article is a comprehensive survey about the 5G e-health systems from the cutting-edge enabling technologies to actual deployments, field trials and future challenges. We expect our work can attract more research attentions to accelerate its world-wide implementations, so that more people can share equal-quality medical service in spite of their living places and conditions.

\begin{table*}
	\centering
	\caption{5G e-health vertical KPIs.}
	\label{tab:1}
	\begin{tabular}{cccccccccc}
		\hline
	& \textbf{Requirements} & \textbf{H1A} & \textbf{H1B} & \textbf{H1C} & \textbf{H1D} & \textbf{H2A} & \textbf{H3A} & \textbf{H3B} & \textbf{Unit} \\
		\toprule
		&\textbf{Downlink speed} &60 & 200 & 10 & 10 & 1 & $\backslash$ & $\backslash$ & Mbps\\
		&\textbf{Uplink speed} &60 & 200 & 10 & 10 & 1 & $\backslash$ & $\backslash$ & Mbps\\
		&\textbf{Latency} &25 &25 &200 &200 &3  &1000 &1000 & Msec\\
		&\textbf{Mobility} & 50 & $\backslash$ &160 & 50 & $\backslash$ & 200 & 200 & Km/h\\
		&\textbf{Interactivity} &1 & 1000 & 1 & 1 & $\backslash$ & $\backslash$ & $\backslash$ & Transaction/sec\\
		&\textbf{Area traffic capacity} &60  &200 &10 &10 &100 &1 &1 &Mbit/sec/$m^2$ \\
		\bottomrule
	\end{tabular}
\end{table*}

\section{Verticals and Enabling Technologies of 5G e-Health Systems}

In this section, we first introduce the verticals and key performance indicators (KPIs) of 5G e-health systems. We afterwards elaborate the 5G enabling technologies of 5G e-health systems.
\subsection{Verticals and KPIs of 5G e-Health Systems}

\begin{figure}[t]
	\centering
	\includegraphics[width=3.5in]{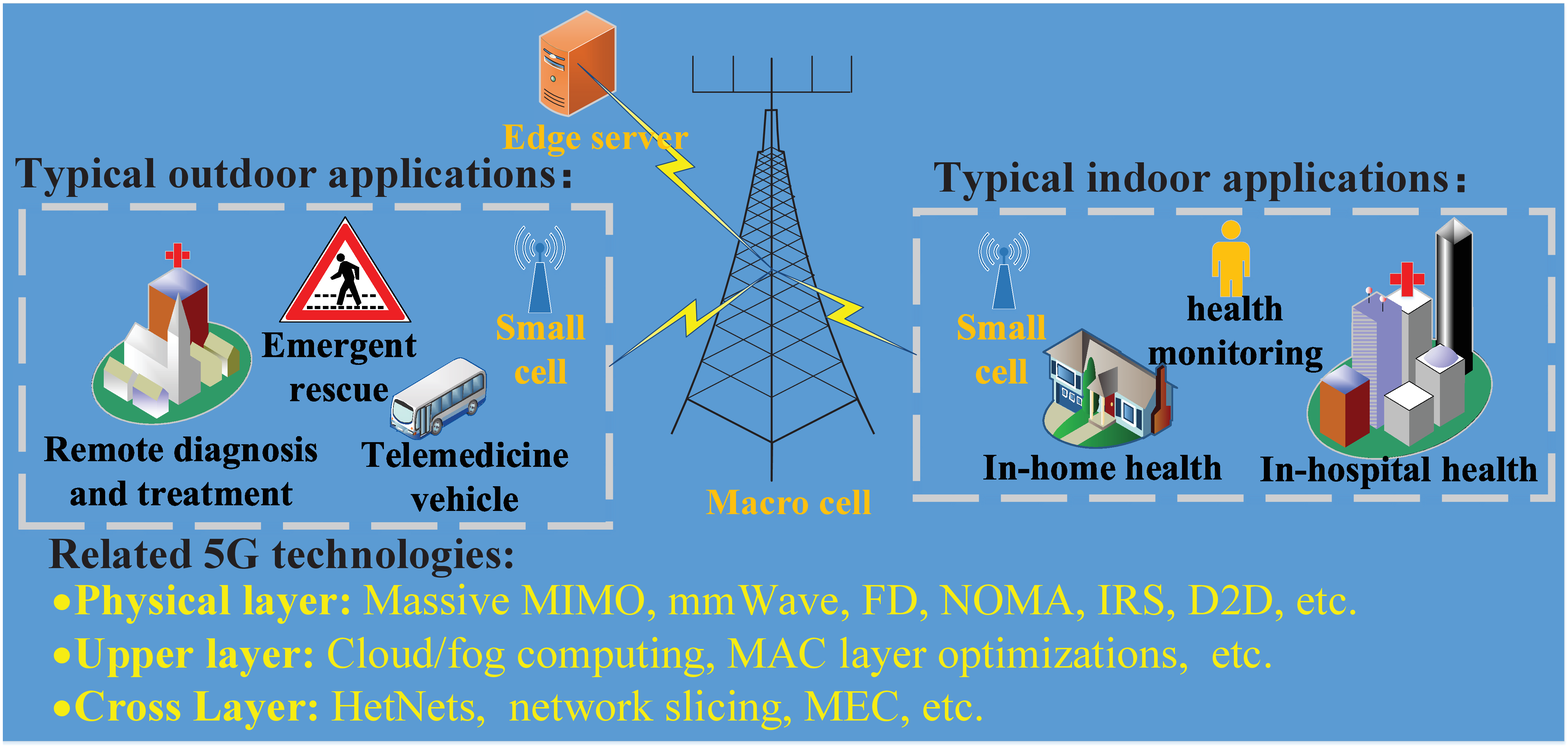}
	\caption{5G e-health applications and related technologies.}
	\label{fig:5gehealth}
\end{figure}

As depicted in Fig. \ref{fig:5gehealth}, health monitoring, emergency rescue are typical outdoor applications of 5G e-health systems, whereas in-hospital e-health (e.g., e-ward-round), remote health are the widely deployed indoor applications. The latest e-health vertical requirements are given in Table \ref{tab:1}, and their related acronyms are listed as follows\footnote{ In line with the European horizon 2020 project \enquote{5G-HEART}, https://5gheart.org/}.
\begin{itemize}
	\item H1A, \enquote{educational surgery}, H1B, \enquote{remote ultrasound examination}, H1C, \enquote{paramedic support}, H1D, \enquote{critical health event}. 
	
	\item H2A, \enquote{the pillcam}, which aims to test real-time transmission with feedback control of a pill camera (capsule video endoscopy) in order to improve the diagnosis effects.
	
	\item H3A, \enquote{vital-sign patch prototype}, H3B, \enquote{localizable tag}. They use vital-sign patches with advanced geo-localisation to explore direct-to-cloud disposable vital-sign patches to enable continuous monitoring of ambulatory patients at anytime and anywhere.
\end{itemize}

As we can see from Table \ref{tab:1}, most of the KPIs are beyond the reach of prior generations of wireless communications. Besides, it is also unpractical and unnecessary to connect all medical devices with optical fiber access networks. For instance, 24-hours health monitoring service greatly relies on the wireless connections but not the optical-based wired connections. Furthermore, we cannot immediately establish a optical-based e-health system while encountering some emergencies. Because of all these reasons, 5G becomes an ideal option for e-health systems.

\subsection{Enabling 5G Technologies for e-Health Systems}

\subsubsection{Physical Layer Technologies for 5G e-Health Systems}

Massive multi-input multi-output (massive MIMO), millimeter wave (mmWave), full duplex (FD), non-orthogonal multiple access (NOMA) and intelligent reflecting surface (IRS) are symbolic 5G physical layer technologies. They can increase the spectrum and energy efficiencies (SEEs) of 5G e-health systems. This is achieved by scaling up the transceivers, allocating wider carrier bandwidth, enabling the simultaneous transmission/reception, sharing the same carrier frequency among multiple users and twisting the wireless channels with highly controllable and intelligent signal reflections. FD and NOMA also can reduce the latency and improve the throughput performance\cite{dzhangtcom}. With the aid of beamforming and directional antenna technologies, we may further extend the coverage area and improve the transmission speed of 5G e-health systems. Besides, device-to-device communications (D2D) can create some localized direct-communication area with or without the help of base station (BS) to offload the wireless access network (WAN) traffic load, and to reduce the transmission latency. Recently, sparse vector coding (SVC)-based superposition is introduced to 5G e-health systems\cite{xwzhang}, which brings in lower block error rate  (BLER) performance while maintaining a massive number of connected health monitoring devices.

However, we cannot always get higher throughput or faster transmission speed simply by allocating wider carrier bandwidth. According to Shannon theory, carrier bandwidth parameter is involved at both the multiplier and the denominator sides, the benefit of wider carrier bandwidth thus becomes marginal as its value growing large. Due to blocking and path-loss effects, ultra-dense deployment (e.g., small/micro cell) is inevitable while implementing the massive MIMO and wider-bandwidth-based 5G e-health systems\cite{busari}. In addition, NOMA receiver needs to employ some sophisticated decoding mechanisms, such as successive interference cancellation (SIC). In this case, although 3rd generation partnership project (3GPP) has adopted NOMA as an optional uplink transmission scheme in 5G, it is still not a matured option, subsequent study on advanced decoding algorithm with low complexity will be needed\cite{xwzhang}.

\subsubsection{Upper Layer Technologies for 5G e-Health Systems}

Cloud computing is one of the widely used upper layer technologies in 5G e-health systems. While applying, we can also invoke the big data, artificial intelligence (AI), and other state-of-the-art technologies to analyse the acquired data for precise diagnosis and treatment\cite{ww}. However, as a centralized computing technology, cloud computing might result in heavy burden to the core network. In order to remedy this, decentralized technologies such as distributed computing, fog (or edge) computing, are exploited to enable privilege access to entry point of the whole networking area to compute, store, communicate, and process data\cite{xhpeng}.

Apart from the widely used centralized and decentralized computing technologies in 5G e-health systems, nowadays increasing studies are on the medium access control (MAC) layer optimizations. Representative technologies are the age of information (AoI) and grant-free transmission (GFT), which are applied to the time-sensitive 5G e-health applications, for example, remote surgery and emergency rescue\cite{szhang}. In literature, AoI aims to measure and optimize the information updates in a network to reduce the consumed time in terms of queuing and scheduling. GFT, on the other hand, enables user to immediately transmit data using the nearest pre-configured resources. Furthermore, AI-based network activity learning and decision making strategies can be adopted for different vertical requirements and appropriate performance requirements of complex mobility scenarios. 

\subsubsection{Cross Layer Technologies for 5G e-Health Systems}

A cooperative architecture from physical, network, MAC and application layers is a prerequisite for cross layer technologies in 5G e-health systems. Among cross layer technologies, network slicing that can create different isolated end-to-end (E2E) slices\cite{Nakao} for medical verticals with shared physical infrastructures is an indispensable one. Network slicing can be divided into WAN slicing, service provider network (SPN) slicing and the core network (CN) slicing parts. While implementing, G.mtn (SPN technology proposed for 5G transport and approved by the international telecommunication union's telecommunication standardization sector (ITU-T), which is compatible with the Ethernet ecosystem and based on slicing the Ethernet core) and flexible Enternet (FlexE) are used to create SPN slices. Software-defined networking (SDN) and network functions virtualization (NFV) technologies are used to create, isolate, and recycle the CN slices from the core network. Besides, wireless caching and mobile edge computing (MEC) proactively cache the hot contents for subsequent sharing to reduce the latency and network burden, which is also widely used in WAN slices. 

In general, 5G e-health cannot be solely achieved by any technology mentioned above, a joint cross layer force is a better choice. For instance, mmWave-massive-MIMO opens up a new solution for 5G e-health systems\cite{busari}, and the full duplex-NOMA-based on a decentralized network architecture has better performance in terms of capacity and latency while compared to other schemes\cite{dzhangtcom}. Machine learning framework with both model-based method and data-based method can be applied in 5G network to estimate the channel condition with or without prior information, which yields better system feasibility\cite{hmwang}. Recently, intelligent communications is emerging as another indispensable topic of 5G e-health systems. It enables the network to predict the forthcoming activities from previous activities, and to allocate the eligible network resources according to the needs of these forthcoming activities.

\section{Use Cases and Field-Trials of 5G e-Health Systems}

In this section, we first elaborate two typical use cases in line with our implementations. We afterwards introduce our field trials and the main test results of our implementations. 

\subsection{5G e-Health Systems for Remote Health}

\begin{figure}[htb]
	\centering
	\includegraphics[width=3.5in]{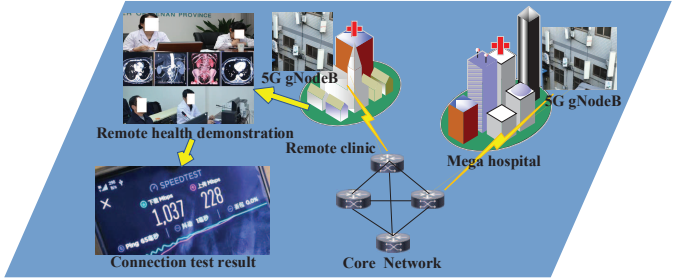}
	\caption{5G e-health systems for remote health.}
	\label{fig:5grh}
\end{figure}

\begin{figure}[t]
	\centering
	\includegraphics[width=3.5in]{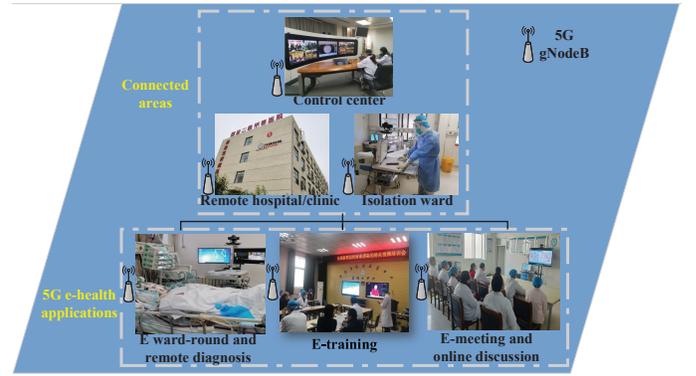}
	\caption{5G e-health systems for Covid-19 pandemic containment.}
	\label{fig:cov19}
\end{figure}

Fig. \ref{fig:5grh} is our implementation of 5G e-health systems for remote health. Statistics show that there are 96.4 million residents in Henan province, and about half of them living in remote or mountainous areas with limited high-quality medical infrastructures (according to the statistical bureau of Henan province, 2018). For them, seeing a doctor is a challenge not only for the money cost but also for the long-distance traveling expenses. In order to provide them some high-quality medical infrastructures, we have connected most of their community clinics and prefecture-level hospitals to the center hub located in the first affiliated hospital of Zhengzhou university (FAHZZU) with 5G wireless networks. In this case, patients from these remote or mountainous areas do not need to go to the capital city to see a doctor while enjoying the equal quality medical service as in FAHZZU.

In our implementation of 5G e-health systems for remote health, high-definition (HD) medical videos and images are transmitted among remote hospital/clinic, hub hospital and mega hospital. Medical experts from these hub and mega hospitals thus can remotely join in the diagnosis, and provide professional suggestions, diagnoses and treatments for patients in remote clinics, as in Fig. \ref{fig:5grh}. However, due to the technical bottlenecks and potential risks of URLLC (as shown in this figure, the latency is much higher than the claimed 1 ms\footnote{Here we adopted the 5G NSA deployment, more details about the current 5G field trial results can be found in section \ref{ft}.}), we didn't demonstrate the critical remote health applications of 5G e-health systems, such as remote surgery. 

\begin{table*}
	\centering
	\caption{Field-trial results of e-health systems with different deployments.}
	\label{tab:2}
	\begin{tabular}{cccccccccc}
		\hline
		\textbf{Scene} &\textbf{Deployment type} &\textbf{Uplink speed} (Mbps) & \textbf{Downlink speed} (Mbps) & \textbf{Latency} (ms) \\
		\toprule
		\multirow{5}{*}{\textbf{Static}} & \textbf{NSA}       &87.00    &595.00   &12.00 \\
		&\textbf{Shared 5G SA}        &94.07   &755.66  &10.22 \\
		&\textbf{Private 5G SA}       &94.15   &767.50    &9.17 \\
		&\textbf{Shared 5G SA-NC}     &89.00   &286.45   &9.24 \\
		&\textbf{Private 5G SA-NC}    &89.65   &772.56  &9.14 \\ \hline
		\multirow{4}{*}{\textbf{Moving}} &\textbf{4G LTE}        &9.93    &21.46   &23.48 \\
		&\textbf{5G NSA}       &87.55   &595.24  &12.00 \\
		&\textbf{Shared 5G SA-NC}  &93.14   &40.60    &8.13 \\
		&\textbf{Private 5G SA-NC} &92.11   &483.10   &8.86 \\
		\bottomrule
	\end{tabular}
\end{table*}

\begin{figure*}
	\centering
	\includegraphics[width=5in]{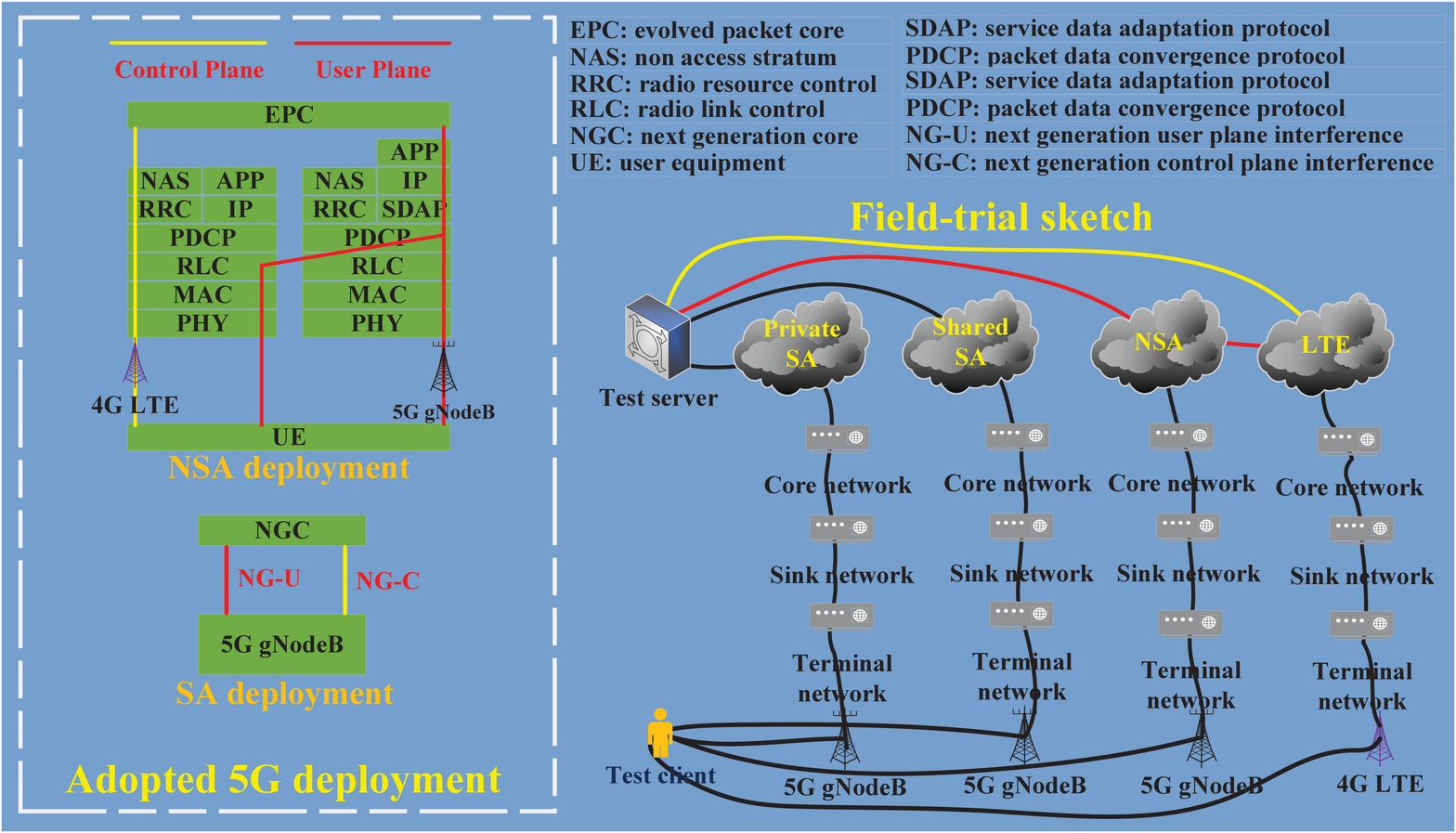}
	\caption{Field trials of 5G e-health systems with 4G LTE, NSA and SA deployments.}
	\label{fig:4g5g}
\end{figure*}

\subsection{5G e-Health Systems for Covid-19 Pandemic Containment}

It is widely agreed that reducing physical contract is needed to block the spread of Covid-19. Staying at home is adopted world widely as a common sense. Most of the schools, stores and entertainment places are closed during this pandemic. It might be easy for ordinary people to keep the social distance, yet how to reduce the physical contract between medical staff and the infected citizen is a challenge. Thanks to 5G e-health systems, we can combine the vital-sign sensors with connectivity to other monitoring devices to remotely monitor the infected citizens. In this case, physical contacts between medical staffs and infected citizens are greatly reduced. Additionally, 5G e-health systems also enables the communication capacity from remote negative pressure isolation wards, prefecture-level hospitals, communities, mega hospitals and municipal control centers to accelerate the information exchanges about Covid-19, so that correct and effective containment methods can be immediately and widely adopted.

From the first day Wuhan was closed (January 29), we activated our 5G e-health systems, and connected more prefecture-level hospitals, community clinics to the mega hospital, FAHZZU, with 5G wireless networks. Thanks to the 5G e-health systems, infected citizens could be remotely monitored and diagnosed by the respiratory and pneumology medical experts from our hub and mega hospitals. In our implementation case, 5G e-health systems provided a direct communication channel for the medical doctors and their infected patients with reduced physical contacts, and it helped the medical doctor to timely diagnose a large number of infected citizens from different places. It also offered e-training and online discussion functions to expedite the latest information flow about Covid-19. We traced the infected citizens in Henan from January 27, 2020 to February 05, 2020 during this Covid-19 pandemic. There were 161 infected cases including 38 severe cases were remotely treated and diagnosed. With the help of 5G e-health systems, although infected cases grew quickly, severe cases and passed away cases were contained.

In addition to the 5G e-health systems, hierarchical diagnosis and treatment is another indispensable method to contain the Covid-19 pandemic. In Henan province, we left the mild cases to medical doctors of the communities, clinics, and prefecture-level hospitals, whereas medical experts from the hub and mega hospitals were devoted to remotely monitor and offer actual treatment for the severe cases if needed. According to our experience, giving enough isolation wards, the pandemic can be well contained. However, if the test-positive citizen number exceeded that of isolation ward, outbreak subsequently came. 

\subsection{Field Trials of 5G e-Health Systems}
\label{ft}

In order to evaluate the technical effectiveness of 5G technologies in e-health systems, we employed a series of field trials to compare the 4G LTE, 5G NSA and 5G SA deployments. As depicted in Fig. \ref{fig:4g5g}, 5G NSA adopted the option-3x, and 5G SA used the option-2 in line with 3GPP Release-15. We first divided the field trials into static scene and moving scene. In the static scene, distance between BS and user terminal is about 50 m, and the moving speed of user terminal is 30 Km/s in the moving scene.  In addition to that, NC was to denote the network congestion, and 4.9 GHz carrier frequency with a total 1 GB carrier bandwidth was employed in the field trials. 

Field trial results of our 5G e-health systems are given in Table \ref{tab:2}. As shown here, the performances of 5G in terms of transmission speed and latency outperform the 4G LTE's. Additionally, downlink speed of the moving scene is slightly lower than that with static scene, whereas the uplink speeds are almost of the same value. On the other hand, while encountering NC, downlink speed of the shared SA is greatly reduced, whereas NC has less effect to the private SA deployment. In this case, dedicated privacy SA might be needed while implementing the 5G e-health systems. Moreover, as we can see from this table, the claimed less than 1 ms latency still cannot be achieved by the current 5G technologies. Subsequent studies and optimizations are still needed in the forthcoming evolution of 5G.

\section{Future Research Trends and Challenges}

Prior discussions demonstrate that the current 5G technologies still cannot fully sustain the e-health systems, especially for the URLLC related applications. Besides, medical application, ethic issue, and international standardization, are some future challenges while deploying the 5G e-health systems.

\subsection{URLLC}

As mentioned in Fig. \ref{fig:5grh} and Table. \ref{tab:2}, latency ability of the current 5G technologies is still far away from the claimed less than 1 ms latency requirement of URLLC applications. In addition to that, less studies have been done on the communication reliability. Although re-transmission and GFT can enhance the communication reliability, it will bring in higher transmission latency performance. On the contrary, applications of 5G e-health systems sometimes rely on both ultra-reliable and low latency wireless connections. 

For future studies, some methods that can improve both latency and reliability performances will be needed. To this end, a combination of multiple technologies from a cross layer perspective might be a feasible solution. For instance, in order to achieve performance targets of low latency and high reliability communications, combining the physical layer technologies (e.g., SVC, mmWave) and upper layer technologies (e.g., AoI, customized network slicing, MEC), and some other emerging technologies might be a practicable method.

\subsection{Medical Applications}

In literature, image processing-based precision medicine technology is adopted to remedy the weak-points of human eyes on detecting the invisible negative lesions. Medical image processing technologies also enable the doctor to trace-back, compare and forecast potential diseases. Due to large scale and even fast growing data volume, effective and fast medical image processing algorithms are required to reduce the process complexity, and to enhance the accuracy in terms of detection and forecast. To cope with these trends, endeavors on high capacity processing chips will be needed in the future.

Precious medical treatment with 5G e-health systems relies on the multi-sources heterogeneous health and clinic data fusing and mining technologies. Medical data fusing and mining ask us to identify the specific fusion level, and extract the feature information. This is a challenging task due to the uncertainly, incompleteness and unstable features of multi-source medical data. Current solutions such as information processing and estimation, statistical inference, decision theory, etc., are incapable to cope with the massive and even growing trends of medical data. For future studies, big data and artificial intelligence-assisted processing methods that can process the large volume data more effectively will be some interesting topics. On the other hand, molecular communication technology that investigates the nano-scale molecular communication mechanisms, is emerging as another hot topic of 5G e-health applications. We may also invoke the molecular communication technology for precious targeted drug delivery, health monitoring, regenerative medicine and genetic engineering\cite{ydeng}.

\subsection{Ethical Challenges}

While sharing the high-quality medical resources to achieve individual opportunities and exercise individual autonomy with 5G e-health systems, fairness legitimacy and criterion are needed. It is the social obligation to offer open discussion, technical training and education with real information about the 5G e-health systems. For instance, according to the European convention on human rights (ECHR), personal data is a distinct right and needs to be controlled over and be free from intrusion into the person's life\cite{more}. In future, while using the 5G e-health systems for e-diagnosis and e-treatment, patient's dignity and autonomy need to be considered for the storage, access and share of the health information. 

Personal information and health data are collected and transmitted with the massive connected smart devices in 5G e-health systems. The data might be leaked out or eavesdropped by malicious hack attacks during the collection and transmission periods. Information protection thus is emerging as a significant issue of 5G e-health systems. Previous studies on secured transmission and information protection greatly rely on the passive exogenous security methods. Therefore, attackers can always update its attack tools or bypass these protective methods. Different from these passive exogenous security methods, endogenous security uses the intrinsic property of wireless communications to enhance the transmission security, which can defense the escalating attack methods. In endogenous security studies, physical layer technologies such as adaptive beamforming, fingerprint-based transceiver classification and verification, upper layer technologies such as mimicry security and AI-based physical layer authentication, are emerging topics for future studies.

\subsection{International Standardization}

International standard has a key role for interoperability, efficiency and safety issues of 5G e-health systems across nations. Global consensus on 5G e-health systems from both design and implementation perspectives needs to be reached first before its world widely deployment. As we know that, adoption of 5G e-health systems requires extensive sharing of information from hospitals to hospitals, organizations to organizations. However, conditions vary from country to country, region to region, which makes the uniform standard across nations a challenging issue. Secondly, legitimacy and ethical law's diversities from places to places, geographical and landscape differences across countries, are making the international standardization of 5G e-health systems an even challenging issue. For future studies, international agreements on unite efforts from different international associations (e.g., ITU, 3GPP) might be required, and the international standardization actions should recognize the differences across nations and regions. 

\section{Conclusion}

This article is a humble attempt to provide a general scope on design and implementation of 5G e-health systems. We discussed the 5G technical solutions for e-health systems, elaborated two use cases according to our implementations, i.e., 5G e-health systems for remote health and 5G e-health systems for Covid-19 pandemic containment. We employed some field trials to compare the network performance among 4G LTE, 5G NSA, 5G shared SA and 5G private SA, and found that substantial studies on URLLC are still needed. We finally outlined the medical applications, ethical and international standardization challenges about the 5G e-health systems.

\begin{bibliographystyle}{IEEEtran}
	\begin{bibliography}{IEEEabrv,bibtex}
	\end{bibliography}
\end{bibliographystyle}

\end{document}